# Evolution of Target Localization in Wireless Sensor Network (WSN): A Review

Muneeb A. Khan[1,2], Muazzam A. Khan[1], Maha Driss[3,4], Wadii Boulila[3,4], Jawad Ahmad[5]
[1]Department of Computer Science, Quaid i Azam Univeristy, Islamabad, Pakistan.
[2]Department of Software, Sangmyung University, South Korea.
[3]RIADI Laboratory, University of Manouba, Manouba 2010, Tunisia.
[4]College of Computer Science and Engineering, Taibah University, Medina 42353, Saudi Arabia.
[5]School of Computing, Edinburgh Napier University, United Kingdom.

*Abstract*—Wireless Sensor Network holds a pivotal position and gained a lot of attention from researchers in recent years. Sensor nodes have been used in vast applications such as environment monitoring, security purpose applications and target tracking. This latter comprises of detection and monitoring of the target movement. In this paper, we explore in detail well-known target tracking techniques. The existing techniques are evaluated using metrics such as network topology, target recovery, energy efficiency and security. We also discuss some of the challenges that affect the performance of tracking schemes. Furthermore, a thorough analysis is performed on existing techniques and future directions are explored.

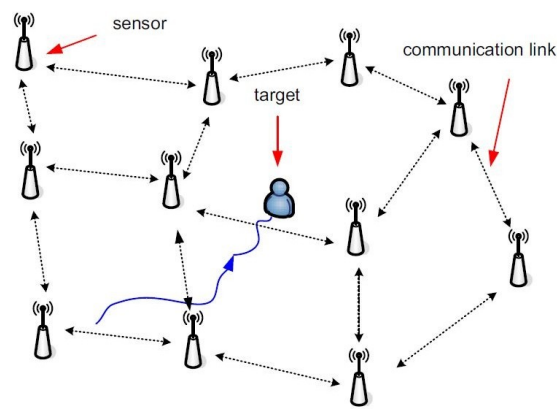

Fig. 1. Target Tracking in WSN

## I. INTRODUCTION

Wireless Sensor Network (WSN) is very essential in the development of a smart environment; e.g. smart cities, smart building, smart grid, transportation and shipping system, etc. It acts as a mediator between the real world and smart systems. WSN is comprised of small devices, called sensors node (SN), ranges from few to hundreds of thousands [1], [2]. An SN varies in size from a grain of dust to shoe size. The cost of SNs similarly varies from a few cents to hundreds of dollars depending on the complexity of each node. SNs perform multiple tasks such as monitoring/sensing, processing, gathering information, and communication [3]. Due to their low price and independent nature with human interference, they are used in different areas for monitoring like ambient monitoring [4], [11], health monitoring [5], [6], underground and underwater systems [7], industrial equipment, surveillance [8]–[10]. Distributed SNs interconnect with each other in such a manner that they act as a single unit. SNs sense or acquire a specific type of data and forward the information to the sink node or cluster head (CH) depending on the topology. SNs perform multiple tasks such as monitoring/sensing, processing, gathering information, and communication [3]. The SNs have small batteries, which are mostly irreplaceable and not chargeable so efficient energy consumption is among the critical issues for SNs.

SNs are also used to track and report the positions of moving objects. Target tracking (shown in Fig. 1) is one of the most important applications in WSN in which SNs keep track of the target and report the location to the user's application. Target tracking can be used in many fields such as campus monitoring, habitat monitoring, health monitoring illegal border crossing, and battlefield surveillance.

However, there are various challenges such as energy efficiency, accuracy, forwarding strategies, load-balancing, prediction and recovery, to overcome for a reliable and efficient target tracking [43]–[45]. For example, Health-related applications need fast and reliable data transfer while energy-efficient and robust methods are required for monitoring applications in smart environments. In addition, costs, hardware selection and connectivity are also a challenge for efficient tracking.

Tracking can be done either with the help of a single SN or by the collaboration of multiple SNs. The use of a single SN causes rapid depletion of energy, extensive computation, and low accuracy. However, multiple SNs give better accuracy, more energy efficiency, and less computation as compared to a single SN.

Multiple classifications of target tracking have been proposed in the literature. In this paper, we have studied well-known tracking algorithms according to security aspects [4], efficient energy consumption [5], [8], node clustering [28], and precision [12].

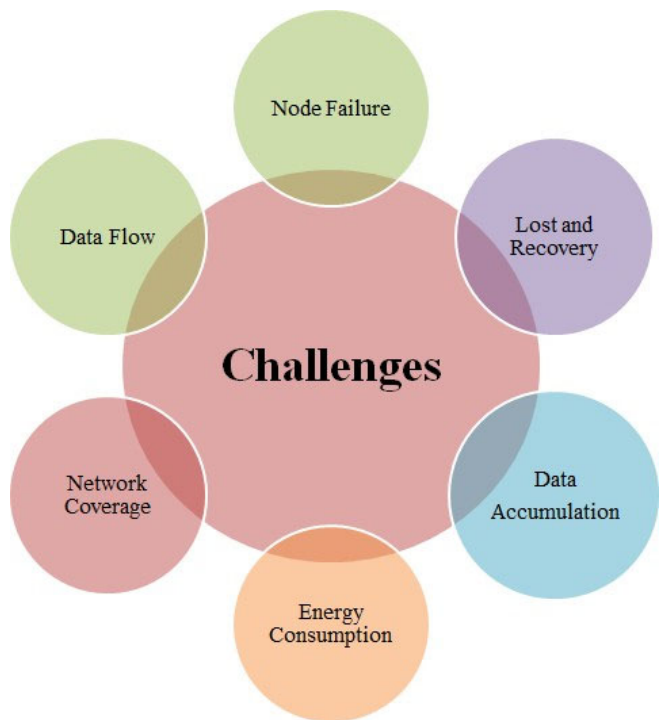

Fig. 2. Challenges of Target tracking

## II. CHALLENGES IN LOCALIZATION AND TARGET TRACKING

Many challenges are faced during the localization and target tracking as shown in Figure 2. These obstacles affect the overall tracking efficiency of WSN such as:

- **Node Failure**: In WSN, SNs are liable to failure due to battery depletion, the occurrence of catastrophe, hardware failure, and external attack. This justifies the need for protocols that can cope up with these challenges.
- **Lost and Recovery of Target**: Prediction errors, hurdles in the network, change of path or speed causes the loss of target. A robust tracking algorithm is highly needed to tackle this problem.
- **Data Accumulation**: In cluster based tracking, SNs forward their data packets to their associated CH. After receiving these packets from SNs, CH accumulates the data and accurately removes the repetition and duplication. During this process of accumulation, data latency and energy consumption are tried to keep minimum as much as can.
- **Energy Consumption**: SNs run on batteries that are non-rechargeable and sometimes in a non-changeable environment. Due to which energy efficiency is a very serious issue in WSN specifically in sensitive target tracking applications. Such energy efficient algorithms are needed to resolve this issue and prolong the network lifetime.
- **Network Coverage**: Target tracking and coverage of network area are directly related and work for hand in hand. The overall performance of the tracking algorithm depends on the coverage of the network. If the network contains holes or SNs are scantly distributes, the performance of the network is degraded. For better accuracy from target tracking, the network must not have any holes or sparse.
- **Data Flow**: The need for data flow change from situation to situation like normal scenario or emergency scenario. In case of emergency or mission-critical applications delay and interruption affect the tracking efficiency. So, such schemes should be designed in which there is no or minimal interruption.

## III. CLASSIFICATION OF TARGET TRACKING APPROACHES

In the literature review, different angles of target tracking are studied. Target tracking can be characterized into multiple aspects: efficient energy consumption, network structure, precision, target loss and recovery, and so on.

In this section, we discuss the taxonomy of target tracking and metrics according to which we assess the recent algorithm. Fig. 3 shows the possible taxonomy of target tracking WSNs.

### A. Network Structure

In [10], the author has distributed the network structure into three categories: Tree based structure, cluster, and leader based network structure. They have coupled each category with prediction based strategies to make protocol more energy efficient and minimize chances of target loss. In [13], the author divides the network structure into two types: hierarchical and peer to peer. The hierarchical network further divided into four structures: activation based, cluster based, tree based and hybrid. The peer to peer based network composed of embedded filters: distributed Kalman filtering (DKF) and distributed scalable Sigma-Point Kalman filter (DSPKF).

We differentiate network structures into three types: Flat structure, Tree based and Cluster based network structure.

- *Flat Structure* It is a type of no topology or the lack of topology. In a flat network, each SN contributes an equal role in network development and establishment. In this topology, SNs broadcast data until it reached the destination SN. This architecture doesn't consider the efficient energy consumption of SNs.
- *Tree Structure* In tree based network structure, deployed SNs construct logical tree based architecture. Data travel from leaf SN to root SN. This way energy is preserved because it reserves SNs from packet flooding and broadcasting.
- *Cluster Structure* Structure is shaped when the network is deployed and its responsibilities are defined such as the number of SNs, coverage area etc. Cluster based topology provides us with scalability and efficient use of bandwidth than other topologies. If CH is selected through local network processing fewer packets are transmitted to a base station which results in efficient energy consumption, efficient bandwidth usage, and security. Clustering can be static or dynamic.

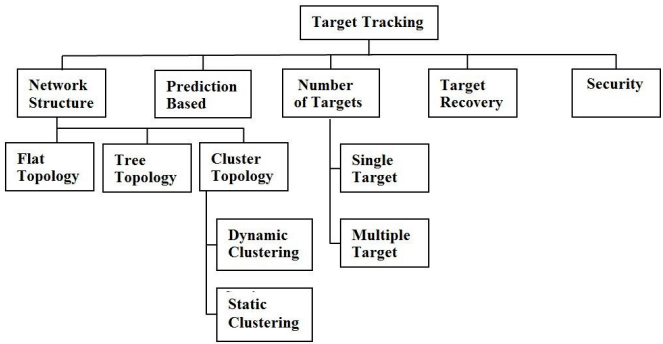

Fig. 3. Taxonomy of Target tracking

1) *Static Cluster*: are formed during network setup and remain constant throughout the network's lifespan. Apart from its convenience, it has many disadvantages, including the fact that the entire cluster's existence is dependent on CH, and there is no likelihood of data sharing or cooperation among clusters.
2) *Dynamic Cluster*: The clusters from dynamically as the target moves. It is better compared to static clustering due to its stability. As a result of the need, new clusters emerge. As soon as the target is entered, the cluster is formed, while the other SNs remain asleep. Just one cluster is enabled at a time, resulting in energy conservation; however, the flaw in dynamic clustering is data redundancy and interfering issues.

*B. Prediction Based Tracking*

Prediction techniques are used in prediction-based tracking methods to predict the future target position. SNs are enabled near the next location, while others stay asleep, saving the energy of SNs and enhancing the total life time of the network. Different techniques and models are suggested to predict the next expected position of the mobile target such as Kalman filter (KF), extended Kalman filter (EKF) [29], [30], linear prediction model [17], Unscented Kalman Filter (UKF) [40] etc.

*C. Number of Targets*

Target tracking can be divided into two categories depending on the target: single target tracking and multiple target tracking.
- *Single Target* Target Tracking a single target consumes less power and energy efficient. It generates a low traffic load during the target tracking.
- *Multiple Targets* Tracking multiple targets is a difficult task. It becomes more difficult due to the differences in target speeds and directions. Multiple target information is received by SNs. The most difficult challenge is determining which data belongs to which target. The data association problem arises from this improbability of knowledge.

*D. Target Recovery*

Prediction based algorithm in WSN sometimes suffers target loss due to sudden speed change, node failure, computer loss, error in location estimation, prediction error, and network coverage problems. prediction-based algorithm. In order to retrieve the lost target, it is necessary to have a robust recovery process. Rapid, resilient and energy-efficient recovery mechanisms. In this area, a great deal of research has been done to address this problem.

*E. Security*

In target tracking, security is one of the key issues due to certain mission-critical applications. These SNs can be deployed to unfriendly, hostile areas in mission-critique applications, and intruder/enemies can easily compromise or capture these SNs. These affected SNs may cause false, fake data transmission, such as the exact location of the objective, and may make the tracking reliability dubious.

IV. COMPARISON AND ANALYSIS OF EXISTING WIRELESS TRACKING TECHNIQUES

In flat topology, all the protocols are trying their best to find the optimised route from source to sink SN, using some sort of flooding. This broadcasting in the network causes rapid battery depletion, packet overhead and shortens the network lifetime. Some probability based techniques are designed to decrease duplicate packet flooding and to institute routing path like Sensor Protocol for Information via Negotiation (SPIN), Rumor routing etc.

1) **SPIN [18]:** In Sensor Protocol for Information via Negotiation (SPIN) message flooding issue is solved with the help of negotiation. It consists of three types of messages (i) Advertisement (ADV), (ii) Request (REQ), (iii)Data (DA). As soon as, SNs receive some new data, SN sends an ADV message to a neighbor. Interested SN sends REQ message to the sender. As a result, a DATA message is sent to the interested SN. SPIN improve network lifetime by reducing flooding and redundant data. The flaw of SPIN is that it does not guarantee data delivery.
2) **RR [19]:** The author proposes a Rumor Routing (RR) protocol. RR performed only for small WSN. However, in the case of large area WSN the maintenance of agents (SNs) and table in each SN is become very complex. Overhead of RR is related to multiple factors like time to live, number of agents and queries. Leader SN becomes aware of target or event from event agents. Leaders SN after applying the heuristics technique decide the route for the next hop selection. The main advantage of RR is no topology maintenance and good quality of routes, on the other hand, disadvantages of RR is unreliability, high delay, and unawareness of new neighbor arrival and dead neighbor.
3) **RR [20]:** In [20], authors proposed an object tracking algorithm based on the Fuzzy Sensing Model in correspondence with RSSI (Radio Signal Strength Indicator) to track the target. After tracking the target, SN finds its location via GPS and sends the calculated target location to the sink SN. They proposed a tree structure based tracking called "convoy tree", which will be created as soon as the target is tracked. All nodes near the target are connected to the tree. Thus giving us 100% coverage and far nodes will remain in sleep mode, results in saving energy. The proposed algorithm works better in term of energy efficiency and mobile target tracking.

4) **ETX-NH [21]:** The author proposed a novel routing protocol based on Neighborhood Heuristics (NHs) model for tree structured WSN. NH coupled SNs routing metric with its neighbors (like energy, distance) to spotlight available routes. This additional information regarding neighbors assists in the selection of the route and maintains the overall routing quality. The proposed routing scheme is energy efficient and performs better even in a lossy network environment.

5) **DLSTA [22]:** Dynamic Look ahead Spanning Tree Algorithm (DLSTA) is suggested in [22] to minimize the chances of a target lost by pre-constructing look ahead cluster along with the target predicted trajectory before it arrives. The SN closest to the destination is chosen as root SN in DLSTA Root SN constructs the tree cluster; calculate the location, next predicted position, speed. The creation of a pre-constructed tree depends upon the velocity of the target. Multiple filters like extended filters Kalman (EKF) and particle Filter (PF) in DLSTA are used to ensure accurate prediction. This accurate prior knowledge assists in preserving the energy of SN and improves the overall network lifetime.

6) **DHSCA [23]:** A novel Dual Head Static Clustering Algorithm (DHSCA) is proposed in [23] to improve the overall network lifetime and energy consumption. In DHSCA; the network is comprised of static clusters based on of the geophysical location of SNs to eliminate the cluster formation overhead like in dynamic clustering. Two nodes are selected as CH; one for data aggregation called Aggregating Cluster Head (ACH) and the other for data transmission called Transmitting Cluster Head (TCH), based on residual energy and distance from other SNs. DHSCA maintains the balance of energy consumption among SNs and improves the WSN lifetime by selecting dynamically CH.

7) **BCTT [24]:** In [24], a new target tracking algorithm; Boundary Static Clustering Target Tracking (BCTT), is proposed. BCTT solve the boundary tracking issues by allowing the boundary sensor to become a member of as many clusters as they want. BCTT also allow SN to collaborate and share information, by allowing this the overall tracking efficiency is increased. The proposed scheme is better than dynamic and hybrid because there is no overhead of cluster formation and destruction.

8) **SCDCH [25]:** A Static Cluster and Dynamic Cluster Head (SCDCH) algorithm is proposed in [25]. Author coupled SCDCH with Newton-Gaussian algorithm to predict the target trajectory and error estimation. SCDCH is used to collect the data from active SN and forward it to the CH. This protocol gives high accuracy and reduces energy consumption, thus prolong the network lifetime overall.

9) **DCTT-PCTT [26]:** The author proposed two algorithms; (i) Distributed Cluster-based Algorithm for Target Tracking (DCTT) and Prediction-based Clustering algorithm for Target Tracking (PCTT), for vehicular tracking in a Vehicular Ad-hoc Network (VANET). DCTT is a dynamic clustering algorithm in which CH is responsible for target location information, aggregation of data received and forwarding it to sink node called Command and Control Center (CC). A Target Failure Probability (TFP) is maintained and shared by SNs among each other. TFP is a metric designed for the selection of CH. If CH is lost, the best node with the minimum TFP is selected as a CH.
PCTT is a centralized VANET based clustering algorithm. In PCTT, CH is responsible for the management of cluster and target tracking. Permission to join the cluster, calculation and selection criteria for CH, maintenance decisions all are executed by CH.

10) **ACDF [27]:** A dynamic clustering based adaptive filtering scheme is proposed for target tracking in a WSN. A two-stage hierarchical data aggregation technique by keeping in mind energy efficiency. At the first stage, SN calculates their estimated distance from the mobile target and shares it with all cluster members and CH. Apart from receiving the SN estimated distance calculation, CH also calculates its distance from the target and aggregate it with the received data from SNs. CH will be selected based on the of residual energy. The proposed scheme gives us better and accurate target tracking in an energy efficient way.

11) **RSSI-LPM [28]:** A scheme is proposed to track and predict the next location of the target. The network is divided into static clusters. CH will be selected whether based on residual energy or distance from the base station. Trilateration mechanism is used to track the target and unify this tracking mechanism with a Linear Prediction model to predict the next location of the target. Only SN, closest to the next location will be active while the rest will remain asleep. This mechanism maintains high accuracy, improve network lifetime.

12) **ARIMA-UKF [29]:** An energy efficient tracking scheme is proposed by unifying two algorithms; Auto Regressive Integrated Moving Average (ARIMA) and Unscented Kalman Filter (UKF). ARIMA is a time series based statistical method in which after observing the target in equal intervals, its future location is predicted based on its past with the least possible error. UKF gives the estimated target position. This use of ARIMA and UKF preserve the overall energy of SN and improve the network lifetime.

13) **LSA-RCAM [30]:** An accurate and energy efficient

TABLE I
COMPARISON OF EXISTING TARGET TRACKING TECHNIQUES

| Sr# | Methodology | Network Structure | | | | Energy Efficiency | Prediction | Number of Targets | | Target Recovery | Security |
|---|---|---|---|---|---|---|---|---|---|---|---|
| | | *Flat* | *Tree* | *Cluster* | | | | *Single* | *Multiple* | | |
| | | | | *S: Static* | *D: Dynamic* | | | | | | |
| 1 | SPIN [18] | Yes | No | No | No | No | No | Yes | No | No | No |
| 2 | RR [19] | Yes | No | No | No | No | No | Yes | No | No | No |
| 3 | FUZZY TARGET TRACKING [20] | No | Yes | No | No | Yes | No | Yes | No | No | No |
| 4 | ETX-NH [21] | No | Yes | No | No | Yes | No | Yes | No | No | No |
| 5 | DLSTA [22] | No | Yes | No | No | Yes | Yes | Yes | No | No | No |
| 6 | DHSCA [23] | No | No | Yes | No | Yes | No | Yes | No | No | No |
| 7 | BCTT [24] | No | No | Yes | No | No | No | Yes | No | No | No |
| 8 | SCDCH [25] | No | No | Yes | No | No | Yes | Yes | No | No | No |
| 9 | DCTT-PCTT [26] | No | No | Yes | No | No | Yes | Yes | No | No | No |
| 10 | ACDF [27] | No | No | Yes | No | Yes | Yes | Yes | No | No | No |
| 11 | RSSI-LPM [28] | No | No | No | Yes | No | Yes | Yes | No | No | No |
| 12 | ARIMA-UKF [29] | No | No | No | Yes | Yes | No | Yes | No | No | No |
| 13 | LSA-RCAM [30] | No | No | No | No | No | Yes | Yes | No | No | No |
| 14 | DPR [31] | No | No | No | No | Yes | Yes | Yes | No | No | No |
| 15 | SJPDA-PUESRF [32] | No | No | No | No | No | Yes | No | Yes | No | No |
| 16 | ASMT [33] | No | No | Yes | No | No | No | No | Yes | No | No |
| 17 | PPHD-MMA [34] | No | No | No | Yes | No | No | No | Yes | No | No |
| 18 | SSN-AUKF [35] | No | No | No | Yes | No | Yes | Yes | No | Yes | No |
| 19 | VGTR [36] | No | No | No | Yes | No | No | No | No | Yes | No |
| 20 | RM-KF [37] | No | No | Yes | No | No | Yes | Yes | No | Yes | No |
| 21 | GTPM [38] | No | No | No | No | Yes | Yes | No | No | No | Yes |
| 22 | SRPTT [39] | No | No | Yes | No | No | Yes | Yes | No | No | Yes |
| 23 | TDKF [40] | No | No | No | Yes | No | No | Yes | No | No | Yes |

target tracking scheme based on prediction is proposed. The Least Square Algorithm (LSA) is used to track the mobile target location and Random way and Constant Acceleration Model (RCAM) for target mobility parameters like speed, direction, and acceleration. The proposed prediction method has good target accuracy, decrease the number of active SN; the only closeat to the target will be active while the rest will remain asleep, this will improve the overall network lifetime.

14) **DPR [31]:** The author proposed a Dual Prediction based Routing (DPR) algorithm. This algorithm improves network lifetime by keeping most nodes in sleep mode. The estimated next location of the target is predicted two times, (i) at nodes level (ii) at sink level. If the difference between them is under the threshold, the sink is not updated which results in lessen the packet transmission. Target is tracked by sensors through trilateration algorithm then forwards the information to their respective CH called "leader", which aggregate the data and forward the information to sink. The proposed scheme gives better results in term of average energy consumption, network lifetime.

Tracking multiple targets has various applications like border monitoring, battlefield monitoring, surveillance, and air traffic control. The main challenges are multi-target tracking tracks, the precise location of targets, associate the clutter and noisy calculations correctly.

15) **SJPDA-PUESRF [32]:** In [32], to track the multiple targets coordinate turn model is adopted with a non-linear turning rate. Sample based Joint Data Association (SJPDA) is used to associate or discard the obtain calculations with the relative target. A variant of Ensemble Square root Filter called Particle wise Update version of Ensemble Square root Filter (PUESRF) is used due to it does not rely on Gaussian distribution and gives a low root mean square error. This integration of SJPDA with PUESRF results in accurate and precise data association as well as consistent targets tracking.

16) **ASMT [33]:** A novel algorithm is proposed for Augmented Specified Multipurpose Tracking (ASMT). ASMT is a centralized Bayesian algorithm based on simplistic estimates. This information is detected by the SN to CH and forwarded to CH. CH is responsible for data association and the target position and nearby location are determined and forwarded to the sink node by using the Bayesian estimation calculation. ASMT offers high-precision multi-target tracking, less calculation and resolves the issue of the WSN data combination very effectively using location and speed status.

17) **PPHD-MMA [34]:** The author has proposed the Multi-path to Measurement Association approach (MMA) and combines it with the Probability Hypothesis Density using a multi-target Particle Filter (PPHD) in urban areas. The PPHD-MMA system need not be aware of how many targets are so that targets can vary over time. K-mean clustering is used to cluster and to compute at any given time the number of targets. PPHD-MMA calculates the probability path at each stage, so that the probability for all paths is identified, and the path that is most likely selected. Due to the use of PPHD and treating the received targets information as random finite set (RFS) measurement to track association problem is avoided.

18) **SSN-AUKF [35]:** A novel scheme is proposed to recover the lost target with minimum SN wake up. An Adaptive Unscented Kalman Filter (AUKF) algorithm is integrated to enhance the robustness and accuracy of the recovery mechanism. The author proposed scheme is a static clustered cooperative network which consists of some static SN (SSN) and a few mobile SN called mobile node (MN) to track the target. Unscented Kalman Filter (UKF) is used to predict the target next location. Once the target is declared lost due to its varying speed or due to holes in WSN, MN will continue tracking.AUKF fine-tunes the noise covariance matrix to improve the recovery mechanism's accuracy and robustness. This vigorous scheduling of MN and SSN improves tracking probability while consuming less energy than just SSN.

19) **VGTR [36]:** An effective camera sensor based scheme called Virtual Grid Target Recovery (VGTR) is proposed in which only necessary cameras are activated to carry useful information about the mobile target. Cameras taking part in mobile target tracking are elected with dynamism depending on mobile target speed. VGTR divide the whole monitoring area into a virtual grid and horizontal, vertical lines. The intersections of these lines are called Virtual Nodes (VN). Each sensor knows the location of its closest VN. If the target is lost due to a sudden change in direction or speed, to recover the lost mobile target all cameras are activated.

20) **RM-KF [37]:** The author has discussed a fast, energy efficient target recovery scheme named Recovery Mechanism (RM) for static clustered WSN. In the proposed scheme, the author used trilateration for tracking and used Kalman filter (KF) for the prediction of the target next position. The author only considers three reasons for target loss; (i) communication failure, (ii) node failure (iii) change in target speed and the proposed scheme RM is applied for target recovery. It is been observed that RM is quick in terms of time and energy efficieny.

21) **GTPM [38]:** The author proposed a scheme to cater for the security issues in WSN target tracking such as fake node positioning, bogus data flooding. A Game Theory Payoff Matrix (GTPM) was developed to select and maintain a balance between efficient energy consumption and accurate tracking. Because of GTPM analysis of different parameters (like the speed of SN, Energy consumption, next location of the target, etc.) has been shifted from technical to strategy field. The author used multiple prediction algorithms (Linear Regression model and Grey Model). The proposed energy efficient scheme can detect and highlight the fake node positioning or bogus data flooding.

22) **SRPTT [39]:** A Secure and Reliable Prediction-based Target Tracking (SRPTT) algorithm is proposed in [28] to prevent rouge SN from faking its location or flooding the bogus information packets in a WSN. SRPTT is a cluster based network algorithm which uses trust information for mobile target tracking. Each SN maintains a table; called the reputation table, in which the good and bad actions of its neighbors are logged. This logged information is exchanged and this information is used during trust evaluation for CH selection and the credibility of information received from SN. By utilizing the reputation concept, SRPTT maintains a balance between security and mobile target tracking.

23) **TDKF [40]:** A trust based distributed Kalman filtering scheme is proposed for secure, reliable target tracking and to improve pliability against attacks. The clustering technique is implemented by using K-mean clustering, to classify the rogue and trustworthy SN. Only trustworthy SN data is fused or forward to CH. Rogue SN detection and localization are a spin-off of the proposed methodology and it is rigorous to the attacks like a random attack, bogus data and reply attack.

Table 1 recapitulates the classification of the different approaches discussed in the paper.

## V. DISCUSSION

A significant amount of research has been conducted in order to make WSNs smart and energy efficient. A system that consumes a lot of energy is unsuitable for the majority of applications. As a result, energy consumption remains one of the most important challenges in terms of smart environment and target tracking. It is also directly linked to the latency and performance of tracking. The number of nodes and transmission range are two important parameters. The number of SNs engaged in target tracking has a substantial impact on accuracy and energy usage. The more the participation of SNs in localization, the greater the localization accuracy and energy consumption. However, Machine Learning (ML) unified with energy harvesting techniques can be suitable candidate to improve and prolong SN lifetime.

To improve target tracking, different prediction algorithms are implemented. However, these energy-starved methods shorten the lives of SNs. In recent years, researchers have proposed various low duty cycle prediction algorithms in which SNs close to the target remain functioning while others are kept in a sleep state.

In mission critical applications, such as battlefield surveillance and remote health monitoring , security is also a crucial issue. Often, SNs are deployed in a hostile environment, where they are prone to corrupt and exploited by intruders. Some authors purpose the use of cryptography or digital signature based security. Furthermore, it is one of the least research area among this field. However, this area needs attention because when a user uses such applications which reveals its personal information can have dire consequences. Compromising user location for certain services and applications such as the health, industry and defense can be harmful and life-threatening [41], [42].

## VI. CONCLUSION

The rigorous and intrinsic nature of Wireless Sensor Network makes them easily deployable everywhere in the field such as disastrous regions, metropolitan cities, industries, underwater and homes. Among different applications of WSN, such as broader monitoring, wildlife habitat monitoring, shipment monitoring, healthcare, and security surveillance, target tracking is considered as the most important one. It is crucial to track different events at earliest before a big damage and take the corrective measures. The massive research in target tracking and manifold application areas motivated us to thoroughly analyze and compare the performance of existing techniques based on network structure, energy efficiency, prediction accuracy, number of targets, target recovery and security. After this extensive comparative analysis we conclude that existing techniques are lagging behind specifically in tracking multiple targets, target recovery and security. Therefore there is still need to propose such techniques which can address these issues and increase its adaptability.

REFERENCES

[1] M. Safaei, A. S. Ismail, H. Chizari, M. Driss, W. Boulila, S. Asadi, and M. Safaei, "Standalone noise and anomaly detection in wireless sensor networks: A novel time-series and adaptive bayesian-network-based approach," *Software: Practice and Experience*, vol. 50, no. 4, pp. 428–446, 2020.

[2] M. Safaei, S. Asadi, M. Driss, W. Boulila, A. Alsaeedi, H. Chizari, R. Abdullah, and M. Safaei, "A systematic literature review on outlier detection in wireless sensor networks," *Symmetry*, vol. 12, no. 3, p. 328, 2020.

[3] Slavi s̆a Tomic. Target localization and tracking in wireless sensor networks. 2017.

[4] Yashwant Singh, Suman Saha, Urvashi Chugh, and Chhavi Gupta. Distributed event detection in wireless sensor networks for forest fires. In Computer Modelling and Simulation (UKSim), 2013 UKSim 15th International Conference on, pages 634–639. IEEE, 2013.

[5] Zhicheng Dai, Shengming Wang, and Zhonghua Yan. Bshm-wsn: A wireless sensor network for bridge structure health monitoring. In Modelling, Identification & Control (ICMIC), 2012 Proceedings of International Conference on, pages 708–712. IEEE, 2012.

[6] Abdul Saboor, Rizwan Ahmad, Waqas Ahmed, Adnan K Kiani, Yannick Le Moullec,and Muhammad Mahtab Alam. On research challenges in hybrid medium access control protocols for ieee 802.15. 6 wbans. IEEE Sensors Journal, 2018.

[7] Luca Ghelardoni, Alessandro Ghio, and Davide Anguita. Smart underwater wireless sensor networks. In Electrical Electronics Engineers in Israel (IEEEI), 2012 IEEE 27th Convention of, pages 1–5. IEEE, 2012.

[8] Zhong Rongbai and Chen Guohua. Research on major hazard installations monitoring system based on wsn. In Future Computer and Communication (ICFCC), 2010 2nd International Conference on, volume 1, pages V1–741. IEEE, 2010.

[9] Tian He, Sudha Krishnamurthy, John A Stankovic, Tarek Abdelzaher, Liqian Luo, Radu Stoleru, Ting Yan, Lin Gu, Jonathan Hui, and Bruce Krogh. Energy-efficient surveillance system using wireless sensor networks. In Proceedings of the 2nd international conference on Mobile systems, applications, and services, pages 270–283. ACM, 2004.

[10] K. Ramya, K. Praveen Kumar and D. V. Srinivas Rao, "A Survey on Target Tracking Techniques in Wireless Sensor Networks", *International Journal of Computer Science & Engineering Survey*, vol. 3, no. 4, pp. 93-108, 2012.

[11] Khan, Muneeb A.; Saboor, Abdul; Kim, Hyun-chul; Park, Heemin. 2021. "A Systematic Review of Location Aware Schemes in the Internet of Things" Sensors 21, no. 9: 3228. https://doi.org/10.3390/s21093228.

[12] Khan, Muneeb A., Muazzam A. Khan, Anis U. Rahman, Asad Waqar Malik, and Safdar A. Khan. "Exploiting cooperative sensing for accurate target tracking in industrial Internet of things. *International Journal of Distributed Sensor Networks* 15, no. 12 (2019) DOI: 1550147719892203.

[13] A. Oracevic and S. Ozdemir, "A survey of secure target tracking algorithms for wireless sensor networks", *World Congress on Computer Applications and Information Systems (WCCAIS)*, 2014.

[14] O. Demigha, W. Hidouci and T. Ahmed, "On Energy Efficiency in Collaborative Target Tracking in Wireless Sensor Network: A Review", *IEEE Communications Surveys & Tutorials*, vol. 15, no. 3, pp. 1210-1222, 2013.

[15] M. Naderan, M. Dehghan and H. Pedram, "Mobile object tracking techniques in wireless sensor networks", *International Conference on Ultra Modern Telecommunications & Workshops*, 2009.

[16] K. Hazra and B. N. Bhramar Ray, "Target Tracking in Wireless Sensor Network: A Survey", *International Journal of Computer Science and Information Technologies*, vol. 6, no. 4, 2015.

[17] A. Kaswan, K. Nitesh and P. Jana, "Energy efficient path selection for mobile sink and data gathering in wireless sensor networks", *AEU - International Journal of Electronics and Communications*, vol. 73, pp. 110-118, 2017.

[18] J. Kulik, W. Heinzelman and H. Balakrishnan, "Negotiation-Based Protocols for Disseminating Information in Wireless Sensor Networks", *Wireless Networks*, vol. 8, no. 2-3, pp. 169-185, 2002.

[19] T. Semong, S. Anokye, Q. Li and Q. Hu, "Rumor as an Energy-Balancing Multipath Routing Protocol for Wireless Sensor Networks", *International Conference on New Trends in Information and Service Science*, 2009.

[20] S. Bhowmik and C. Giri, "Convoy Tree Based Fuzzy Target Tracking in Wireless Sensor Network", *International Journal of Wireless Information Networks*, vol. 24, no. 4, pp. 476-484, 2017.

[21] D. Delaney, R. Higgs and G. O'Hare, "A Stable Routing Framework for Tree-Based Routing Structures in WSNs", *IEEE Sensors Journal*, vol. 14, no. 10, pp. 3533-3547, 2014.

[22] A. Alaybeyoglu, A. Kantarci and K. Erciyes, "A dynamic lookahead tree based tracking algorithm for wireless sensor networks using particle filtering technique", *Computers & Electrical Engineering*, vol. 40, no. 2, pp. 374-383, 2014.

[23] T. Panag and J. Dhillon, "Dual head static clustering algorithm for wireless sensor networks", *AEU - International Journal of Electronics and Communications*, vol. 88, pp. 148-156, 2018.

[24] S. Rouhani and A. Haghighat, "Boundary static clustering target tracking in wireless sensor networks", *6th International Conference on Computing, Communication and Networking Technologies (ICCCNT)*, 2015.

[25] M. Wahdan, M. Al-Mistarihi and M. Shurman, "Static cluster and dynamic cluster head (SCDCH) adaptive prediction-based algorithm for target tracking in wireless sensor networks", *38th International Convention on Information and Communication Technology, Electronics and Microelectronics (MIPRO)*, 2015.

[26] S. Khakpour, R. Pazzi and K. El-Khatib, "Using clustering for target tracking in vehicular ad hoc networks", *Vehicular Communications*, vol. 9, pp. 83-96, 2017.

[27] H. Zhang, X. Zhou, Z. Wang, H. Yan and J. Sun, "Adaptive Consensus-Based Distributed Target Tracking With Dynamic Cluster in Sensor Networks", *IEEE Transactions on Cybernetics*, pp. 1-12, 2018.

[28] P. Joshi and A. Joshi, "Prediction Based Moving Object Tracking In Wireless Sensor Network", *International Research Journal of Engineering and Technology*, vol. 4, no. 7, 2017.

[29] M. Alishahi, A. Hossein Mohajerzadeh, S. Aslishahi and M. Zabihi, "Adaptive Target Tracking Using Prediction in Wireless Sensor Networks", 2016.

[30] M. Mirsadeghi and A. Mahani, "Low Power Prediction Mechanism for Wsn-based Object Tracking", *Procedia Technology*, vol. 17, pp. 692-698, 2014.

[31] V. Dayan and K. Vijeyakumar, "Target Tracking in Sensor Networks Using Energy Efficient Prediction Based Clustering Algorithm", *Procedia Engineering*, vol. 38, pp. 2070-2076, 2012.

[32] R. Jinan and T. Raveendran, "Particle Filters for Multiple Target Tracking", *Procedia Technology*, vol. 24, pp. 980-987, 2016.

[33] K. Xiao, R. Wang, L. Zhang, J. Li and T. Fun, "ASMT: An augmented state-based multi-target tracking algorithm in wireless sensor networks", *International Journal of Distributed Sensor Networks*, vol. 13, no. 4, p. 155014771770311, 2017.

[34] M. Zhou, J. Zhang and A. Papandreou-Suppappola, "Multiple Target Tracking in Urban Environments", *IEEE Transactions on Signal Processing*, vol. 64, no. 5, pp. 1270-1279, 2016.

[35] H. Qian, P. Fu, B. Li, J. Liu and X. Yuan, "A Novel Loss Recovery and Tracking Scheme for Maneuvering Target in Hybrid WSNs", *Sensors*, vol. 18, no. 2, p. 341, 2018.

[36] J. Amudha and P. Arpita, "Multi-Camera Activation Scheme for Target Tracking with Dynamic Active Camera Group and Virtual Grid-Based

Target Recovery", *Procedia Computer Science*, vol. 58, pp. 241-248, 2015.
[37] S. Patil, A. Gupta and M. Zaveri, "Recovery of Lost Target Using Target Tracking in Event Driven Clustered Wireless Sensor Network", *Journal of Computer Networks and Communications*, vol. 2014, pp. 1-15, 2014.
[38] A. Silva, F. Zhou, E. Pontes, M. Simplicio, R. Aguiar, A. Guelfi and S. Kofuji, "Energy-efficient node position identification through payoff matrix and variability analysis", *Telecommunication Systems*, vol. 65, no. 3, pp. 459-477, 2016.
[39] A. Oracevic, S. Akbas and S. Ozdemir, "Secure and reliable object tracking in wireless sensor networks", *Computers & Security*, vol. 70, pp. 307-318, 2017.
[40] C. Liang, F. Wen and Z. Wang, "Trust-based distributed Kalman filtering for target tracking under malicious cyber attacks", *Information Fusion*, vol. 46, pp. 44-50, 2018.
[41] M. Driss, D. Hasan, W. Boulila, and J. Ahmad, "Microservices in iot security: Current solutions, research challenges, and future directions," *arXiv preprint arXiv:2105.07722*, 2021.
[42] M. A. Khan, M. A. Khan, S. Latif, A. A. Shah, M. U. Rehman, W. Boulila, M. Driss, and J. Ahmad, "Voting classifier-based intrusion detection for iot networks," *arXiv preprint arXiv:2104.10015*, 2021.
[43] A. Petrosino, G. Sciddurlo, G. Grieco, A.A. Shah, G. Piro, L.A. Grieco, and G. Boggia, "Dynamic Management of Forwarding Rules in a T-SDN Architecture with Energy and Bandwidth Constraints", *International Conference on Ad-Hoc Networks and Wireless, Springer*, pp. 3-15, 2020.
[44] A.A. Shah, G. Piro, L.A. Grieco, and G. Boggia, "A review of forwarding strategies in transport software-defined networks", *2020 22nd International Conference on Transparent Optical Networks (ICTON), IEEE*, pp. 1-4, 2020.
[45] A.A. Shah, G. Piro, L.A. Grieco, and G. Boggia, "A review of forwarding strategies in transport software-defined networks", *2020 IEEE/ACM 24th International Symposium on Distributed Simulation and Real Time Applications (DS-RT), IEEE*, pp. 1-4, 2020.